\documentstyle{article} 
\begin{document} 
\title{Paradoxical aspects of the kinetic equations} 
\author{C. Y. Chen} 
\maketitle \centerline{Dept. of Physics, Beijing University of Aeronautics}
\centerline{and Astronautics, Beijing 100083, PRC} 
\centerline{Email: cychen@buaa.edu.cn} 

\begin{abstract} Two paradoxical aspects of the prevailing kinetic equations 
are presented. One is related to the usual understanding of distribution 
function and the other to the usual understanding of the phase space. With 
help of simple counterexamples and direct analyses, involved paradoxes 
manifest themselves. \end{abstract} 

\vskip20pt
When the Boltzmann 
equation, thought of as the first equation of the prevailing kinetic theory, 
came out, strong doubts arose, of which many were related to the fact that 
whereas Newton's equations themselves were time-reversible the Boltzmann 
equation, as a consequence of Newton's equations, was time- 
irreversible\cite{pathria}. As time went by, particularly after the 
``rigorous'' BBGKY theory was formulated in the middle of the last 
century[2$-$8], the philosophical concern of the time-reversal paradox 
gradually faded out, and it was believed that the ultimate understanding of 
the related issues had been completely established, at least in the regime of 
classical mechanics. 

However, relatively recent developments of mathematics and physics seem to 
have brought new elements into the picture. In particular, the studies of 
fractals\cite{mandelbrot,mandelbrot1} reveal that there can, at least in the 
mathematical sense, exist functions that have structures of self-similarity 
even at the infinitesimal level. Such functions are intrinsically 
discontinuous and cannot be described by usual differential apparatuses. 
Along this line, an increasing number of scientists are surmising that if any 
similar structures are found in realistic gases, some of conventional 
concepts in the standard theory need to be revised significantly. 

In connection with this, we wish to present and discuss two relatively 
unknown aspects of the standard framework of kinetic theory. Firstly, it will 
be shown that a realistic gas contains, almost always, a significant amount 
of particles whose distribution function does not keep invariant in the 
six-dimensional phase space and cannot be regarded as a continuous one. 
Secondly, it will be illustrated that there are inherent difficulties in 
formulating the particles that enter and leave an infinitesimal phase volume 
element during an infinitesimal time, which suggests in the sense that the 
phase space is more sophisticated than the customary thought assumes. By 
removing abstractness of the matters, all related paradoxes become 
surprisingly simple and straightforward. We are now convinced that if those 
paradoxes had been unveiled at the very 
berginning, kinetic theory would have been renewed several times. 

The basic core of the classical kinetic theory says that if particles of a 
gas do not interact with each other, the distribution function describing 
them satisfies the collisionless Boltzmann equation 
\begin{equation}\label{collisionless} \frac{\partial f}{\partial t} + {\bf v} 
\cdot \frac{\partial f}{\partial {\bf r}}+\frac{{\bf F}}m \cdot 
\frac{\partial f} {\partial {\bf v}} =0, \end{equation} 
which is, according to well-known textbooks\cite{reif,harris}, equivalent to 
the path-invariance of collisionless distribution function 
\begin{equation}\label{invariance} \left.\frac{df}{dt}\right|_{{\bf 
r}(t),{\bf v}(t)}=0, \end{equation} 
where the subindexes ${\bf r}(t)$ and ${\bf v}(t)$ imply that that the 
differentiation is taken along a particle's path. Conceptually speaking, 
equation (\ref{invariance}) can be interpreted as saying that such a gas is 
``incompressible'' in the phase space. The standard theory further states 
that, if collisions between particles are more than negligible, a certain 
type of collisional operator needs to be introduced to replace the zero term 
in (\ref{collisionless}). After having the collisional operator due to 
Boltzmann himself, equation (\ref{collisionless}) becomes 
\begin{equation} \label{collisional} \frac{\partial f}{\partial t}+{\bf 
v}\cdot \frac{\partial f}{\partial {\bf r}} +\frac {{\bf F}}m 
\cdot\frac{\partial f}{\partial {\bf v}}=\int_{{\bf v}_1,\Omega} [f({\bf 
v}^\prime)f({\bf v}_1^\prime)-f({\bf v})f({\bf v}_1)]u\sigma (\Omega)d\Omega 
d{\bf v}_1, \end{equation} 
where the first term on the right side represents particles entering the unit 
phase volume element and the second term particles leaving the unit phase 
volume element. 

Although the formulation briefed in the last paragraph seems stringent and 
has been accepted unanimously in the community, it is not truly sound. 
Ironically enough, even a simple glance at the form of the Boltzmann equation 
(\ref{collisional}) offers intriguing things to ponder. On the left side 
there is a symmetry between ${\bf r}$ and ${\bf v}$ in terms of 
differentiation operation; whereas on the right side the position vector 
${\bf r}$ serves as an inactive ``parameter'' and all the integral operations 
are performed in the velocity space. (The solid angle $\Omega$ is defined in 
terms of the velocities.) This disparity, while 
seeming a bit too schematic to be 
fully convincing, should serve a motivation for us to investigate the whole 
issue more carefully and more thoroughly. 

We first look at whether or not the collisionless Boltzmann equation 
(\ref{collisionless}) and the path-invariance theorem (\ref{invariance}) make 
sense as they intend to. 

In the mathematical sense, the picture provided by the path-invariance 
is quite clear and simple. Consider a sufficiently small volume element 
\begin{equation} \Delta x \Delta y \Delta z \Delta v_x \Delta v_y \Delta v_z 
\end{equation} 
moving together with a certain particle in the phase space; the theorem 
asserts that the particle density within the element (in number herein) does 
not increase or decrease. In well-known textbooks, the theorem is ``proved'' 
by applying the rigorous Jacobian approach[11$-$13]. Here, for the purpose of 
this paper, we wish to illustrate the theorem and its derivation 
in an intuitive way. Fig.~1 
describes what takes place in the two-dimensional $x$-$v_x$ subspace. As time 
passes from $t=0$ to $t=T$, the particles distributed
inside the rectangle 
$\Delta x \Delta v_x$ in Fig.~1a will be distributed inside the parallelogram 
in Fig.~1b. By excluding all external forces (just for simplicity), it is 
obvious that the area of the parallelogram is equal to that of the rectangle 
and the average particle density within the moving volume element keeps 
invariant. On the understanding that particles distribute continuously within
$\Delta x\Delta v_x$ and the size of $\Delta x\Delta v_x$ can shrink to zero,
the invariance of the average particle density can
certainly be interpreted as the
path-invariance of distribution function. The discussion above, 
though simplified somewhat, reflects the essence of the full theorem 
accurately.

It seems, at this stage, that the validity of the path-invariance 
(\ref{invariance}), together with the picture of incompressible fluid in the 
phase space, is so solid and so clear that we should discuss it no more. But, 
somehow,
the truth is not that simple: there exist many cases in that gases behave 
very differently. To get an immediate idea about such behavior, let's look at 
the special arrangement shown in Fig.~2, where particles with definite 
velocity ${\bf v}$ strike a convex solid surface  and then get reflected 
from it elastically. (Later on, more general models about collisions between 
particles and boundaries will be adopted and examined.) Following a moving 
particle and counting particles in a definite position range $d{\bf r}\equiv 
dxdydz$ and in a definite velocity range $d{\bf v}\equiv dv_x dv_y dv_z$, we 
find out a clear-cut fact that the reflected particles simply obey 
\begin{equation}\label{c1} \left.\frac{df}{dt}\right|_{{\bf r} (t), {\bf 
v}(t)} <0 \quad {\rm and }\quad \frac{\partial f}{\partial t} + {\bf v} \cdot 
\frac{\partial f}{\partial {\bf r}}+\frac{{\bf F}}m \cdot \frac{\partial f} 
{\partial {\bf v}} <0. \end{equation} 
That is to say, these particles diverge in the phase space. In a similar way, 
we can readily see that if the incident particles fall upon a concave solid 
surface, the reflected particles will, in some region, obey 
\begin{equation}\label{c2} \left.\frac{df}{dt}\right|_{{\bf r}(t),{\bf v}(t)} 
>0. \end{equation} 
This tells us that the particles converge. 

A sharp question arises. How can the path-invariance theorem, 
considered as the very core of all the kinetic equations, possibly suffer 
from such simple and direct counterexamples? By reviewing the derivation of 
the Boltzmann equation, we realize that equations (\ref{collisionless}) and 
(\ref{invariance}) hold only for perfectly continuous distribution functions, 
while the diverging and converging particles presented above are related to 
none of them. 

To see the point more vividly, let's go back to the two-dimensional phase 
space, namely the $x$-$v_x$ space, and investigate the evolution of the 
particles marked with the dotted curve in Fig.~2. In the schematic sense, we 
may say that at $t=0$ all these particles are distributed along the diagonal 
of $\Delta x \Delta v_x$ shown in Fig.~3a, rather than in $\Delta x \Delta 
v_x$ uniformly. After a while, at $t=T$, this diagonal is stretched and 
becomes longer as shown in Fig.~3b. If we set up a small, but definite, phase 
volume element and let it move together with one of the particles, we will 
certainly find that the particle density within it decreases. (For  
discontinuous distributions, if we allow the shape and size of measuring
volume element to vary, we can get any value, drastically from 
zero to infinity, for the particle density.) 

Now, it is in order to examine how a realistic boundary surface $S$ 
``reflects'' realistic particles that come from 
all directions and with different speeds. Experimental facts inform us that 
the reflection cannot be truly elastic and stochastic and dissipative 
effects must play a certain role\cite{kogan}. To express particles 
``produced'' by such surface, it is proper (particularly if the gas is a 
rarefied one) to define the instant emission rate $\sigma$ on an 
infinitesimal surface $dS$ in such a way that 
\begin{equation} \sigma dt dS dv d\Omega \end{equation} 
represents the number of particles ejected by the surface element in the 
speed range $dv$ and in the solid angle range $d\Omega$ during the time 
interval $dt$. By allowing the emission rate $\sigma$ to be described in 
probability and to have certain dependence on velocities of incident 
particles, it can be said that the stochastic and dissipative nature of the 
reflection has been included. Then, divide $S$ into $N$ surface elements. 
Referring to Fig.~4, we find that for the $i$th surface element $(\Delta 
S)_i$ the reflected particles are like ones emitted from a ``point particle 
source'' at ${\bf r}_{0i}$, 
the position vector of $(\Delta S)_i$, and the relevant distribution 
function at a point ${\bf r}$ in the reflection region takes the form 
\begin{equation} \label{delta} f({\bf r},v,\Omega)=\frac{\sigma (\Delta S)_i} 
{|{\bf r}-{\bf r}_{0i}|^2 v^3 }V_i(v)\delta (\Omega-\Omega_{{\bf r}-{\bf 
r}_{0i}}), \end{equation} 
where $V_i(v)$ is a certain function of $v$ and $\delta (\Omega-\Omega_{{\bf 
r}- {\bf r}_{0i}})$ is the $\delta$-function defined on the solid angle in 
the velocity space. It is very obvious that, regardless of the forms of 
$\sigma$ and $V(v)$, this distribution function diverges in the phase space. 
Another interesting point about the distribution function is that it is 
perfectly continuous in terms of ${\bf r}$ and $v$ and it is like a function 
defined on a single point in terms of $\Omega$. Based on this observation, we 
are tempted to say that the function is on a variable domain of $4+\epsilon$ 
dimensions. The total distribution function at ${\bf r}$ associated with all 
reflected particles from $S$ is 
\begin{equation} \label{delta1} f({\bf 
r},v,\Omega)=\sum\limits_{i=1}^{N}\frac{\sigma (\Delta S)_i} {|{\bf r}-{\bf 
r}_{0i}|^2 v^3 }V_i(v)\delta (\Omega-\Omega_{{\bf r}-{\bf r}_{0i}}). 
\end{equation} 
Though the distribution function expressed by (\ref{delta1}) can, as 
$N\rightarrow \infty$, be regarded as a sum of an infinitely large number of 
$(4+\epsilon)- $dimensional functions, it is not a continuous function 
defined in the phase space. Actually, with help of this expression, we can 
analytically or numerically prove that the particle number within a given 
moving volume element $d{\bf r} d{\bf v}$ is capable of decreasing, 
increasing or keeping invariant. As a limiting case, we may 
assume $S$ to be 
relatively small, or the distance $|{\bf r}-{\bf r}_{0i}|$ to be relatively 
large, and find that expression (\ref{delta1}) is reduced to expression 
(\ref{delta}) and the particles related to it always diverge in the 
phase space. 

The above investigation, though formally simple, provides us with a very new 
picture on gas dynamics. Rather than as 
a continuous medium or an incompressible 
fluid in the phase space, a gas should be considered as a special collection 
of discrete particles, whose distribution function can change from continuous 
one to discontinuous one, as well as from discontinuous one to continuous 
one. Another notable fact, which may very much interest scientists carrying 
out practical studies, is that the changeovers aforementioned occur  
dramatically near interfaces between fluids and solid boundaries. 

We now turn our attention to the validity of the collisional operator on the 
right side of (\ref{collisional}). A widely accepted concept is that in order 
to formulate collisional effects we are supposed to focus ourselves on a 
fixed six-dimensional phase volume element and study how particles leave and 
enter the element due to collisions. In what follows, it will be shown that 
such concept is, much to our surprise, deceptive. 

Firstly, particles leaving a phase volume element $d{\bf r} d{\bf v}$ during 
a time interval $dt$ are of interest. According to the standard theory, if a 
particle collides with another within the volume element $d{\bf r}d{\bf v}$ 
during $dt$, it should be considered as one that leaves the volume element 
during the time interval because of the velocity change caused by the 
collision (see Fig.~5). This intuitive, seemingly very reasonable, picture 
can be challenged in the following way. In deriving kinetic equations, a 
necessary step is to let $dt$, $d{\bf r}$ and $d{\bf v}$ approach zero 
independently. If it is assumed, in the limiting processes, that the length 
scale $|d{\bf r}|$ 
is much smaller than $|{\bf v}dt|$, then virtually all the particles, 
initially within $d{\bf r}d{\bf v}$, will leave $d{\bf r}d{\bf v}$ at the end 
of $dt$, irrespective of suffering collisions or not. If we still want to say 
that the standard consideration, concerning how many particles stay inside 
$d{\bf r} 
d{\bf v}$ without involving collisions and how many particles get out of 
$d{\bf r} d{\bf v}$ with collisions involved, holds its significance, we have 
no choice but to assume that $|d{\bf r}|>>|{\bf v}dt|$. An unfortunate fact 
is that no sound reason can be found out for that we can prefer this 
assumption to its converse.

Secondly, particles entering a phase volume element $d{\bf r} d{\bf v}$ 
during $dt$ are under examination. In the standard treatment, two beams with 
velocities ${\bf v}^\prime$ and ${\bf v}_1^\prime$ are assumed to collide 
within a volume element $d{\bf r}$ during $dt$ and to give contribution to 
the particles expressed by $f({\bf r},{\bf v},t)d{\bf r}d{\bf v}$, as shown 
in Fig.~6. Ironically enough, this treatment involves not one but many 
paradoxes. As one thing, particles produced by collisions in a small spatial 
region, like ones emitted from a point particle source, will diverge in the 
phase space and cannot be treated as an ordinary contribution to $f(t,{\bf 
r},{\bf v})d{\bf r}d{\bf v}$. As another thing, we again need to let $dt$, 
$d{\bf r}$ and $d{\bf v}$ approach zero. If $|d{\bf r}|<<|{\bf v}dt|$, all 
produced particles will ``instantly'' leave $d{\bf r}d{\bf v}$ and only an 
insignificant fraction of them can be treated as a contribution to $f({\bf 
r},{\bf v},t)d{\bf r}d{\bf v}$. See Ref.~16 to get more paradoxes. 

In summary, two paradoxical aspects of the standard kinetic theory have been 
presented. The first aspect is related to the distribution function. A tacit 
assumption of the existing kinetic equations is that distribution functions, 
though describing discrete particles, must be mentally and practically 
continuous. The discussion of this paper, however, shows that distribution 
functions of realistic gases have, in general, complex local structures and 
they cannot be described by continuous distribution functions. The second 
aspect is related to the phase space. A usual concept in the standard theory 
is that the position space and the velocity space can be separated mentally 
and practically. The discussion of this paper, however, shows that whenever 
we investigate the time dynamics of particles in the velocity space we should 
keep an eye on what takes place in the position space, and vice versa. 

A variety of fundamental questions can be raised, of which many are beyond 
the scope of this brief paper. In some of our recent works, we make more 
analyses and put forward alternative approaches\cite{chen1,chen2}. With help 
of a development in quantum mechanics\cite{chen3}, some of the discussion in 
this paper can also be extended to the regime of quantum statistical physics. 

This paper is supported by School of Science, BUAA, PRC and by Education 
Ministry, PRC.

\newpage
\setlength{\unitlength}{0.018in} 
\hspace{0.8cm}
\begin{picture}(100,100)
\multiput(15,19)(90,0){2}{\vector(1,0){80}} 
\multiput(20,15)(90,0){2}{\vector(0,1){60}} 

\put(45,40){\framebox(20,20){}} 
\multiput(135,40)(20,20){2}{\line(1,0){20}} 
\multiput(135,40)(20,0){2}{\line(1,1){20}} 
\put(50,5){\makebox(20,8)[l]{\bf (a)}}
\put(140,5){\makebox(20,8)[l]{\bf (b)}}
\multiput(89,15)(90,0){2}{\makebox(20,8)[c]{$x$}}
\multiput(10,77)(90,0){2}{\makebox(20,8)[c]{$v_x$}}
\put(46,32){\makebox(20,8)[c]{\small $\Delta x$}}
\put(27,47){\makebox(20,8)[c]{\small $\Delta v_x$}}
\end{picture}
\begin{center}
\begin{minipage}{4.0in}
\vskip-0.2cm
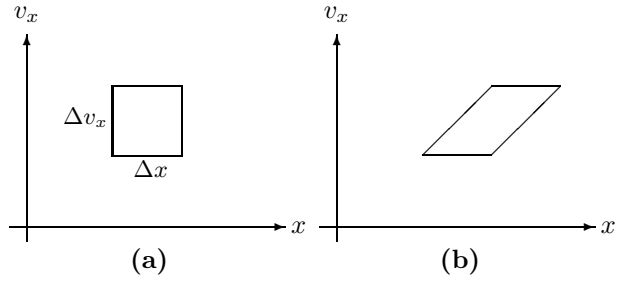
Fig.~1: A moving phase volume element (a) at an initial time $t=0$ and 
(b) at a later time $t=T$.
\end{minipage}

\begin{picture}(100,110)
\hspace{-1.0cm}
\put(30.797, 87.459){\circle*{1.0}}
\put(31.083, 86.931){\circle*{1.0}}
\put(31.369, 86.404){\circle*{1.0}}
\put(31.656, 85.877){\circle*{1.0}}
\put(31.942, 85.349){\circle*{1.0}}
\put(31.942, 85.349){\line(-1,0){5}}
\put(32.228, 84.822){\circle*{1.0}}
\put(32.228, 84.822){\line(-1,0){5}}
\put(32.514, 84.294){\circle*{1.0}}
\put(32.514, 84.294){\line(-1,0){5}}
\put(32.800, 83.767){\circle*{1.0}}
\put(32.800, 83.767){\line(-1,0){5}}
\put(33.044, 83.219){\circle*{1.0}}
\put(33.288, 82.671){\circle*{1.0}}
\put(33.532, 82.122){\circle*{1.0}}
\put(33.776, 81.574){\circle*{1.0}}
\put(34.020, 81.026){\circle*{1.0}}
\put(34.020, 81.026){\line(-1,0){5}}
\put(34.263, 80.478){\circle*{1.0}}
\put(34.263, 80.478){\line(-1,0){5}}
\put(34.507, 79.930){\circle*{1.0}}
\put(34.507, 79.930){\line(-1,0){5}}
\put(34.751, 79.381){\circle*{1.0}}
\put(34.751, 79.381){\line(-1,0){5}}
\put(34.951, 78.816){\circle*{1.0}}
\put(35.152, 78.250){\circle*{1.0}}
\put(35.352, 77.685){\circle*{1.0}}
\put(35.552, 77.119){\circle*{1.0}}
\put(35.753, 76.553){\circle*{1.0}}
\put(35.753, 76.553){\line(-1,0){5}}
\put(35.953, 75.988){\circle*{1.0}}
\put(35.953, 75.988){\line(-1,0){5}}
\put(36.153, 75.422){\circle*{1.0}}
\put(36.153, 75.422){\line(-1,0){5}}
\put(36.318, 74.845){\circle*{1.0}}
\put(36.318, 74.845){\line(-1,0){5}}
\put(36.483, 74.269){\circle*{1.0}}
\put(36.648, 73.692){\circle*{1.0}}
\put(36.813, 73.115){\circle*{1.0}}
\put(36.978, 72.538){\circle*{1.0}}
\put(37.143, 71.961){\circle*{1.0}}
\put(37.143, 71.961){\line(-1,0){5}}
\put(37.308, 71.384){\circle*{1.0}}
\put(37.308, 71.384){\line(-1,0){5}}
\put(37.472, 70.807){\circle*{1.0}}
\put(37.472, 70.807){\line(-1,0){5}}
\put(37.600, 70.221){\circle*{1.0}}
\put(37.600, 70.221){\line(-1,0){5}}
\put(37.728, 69.635){\circle*{1.0}}
\put(37.856, 69.049){\circle*{1.0}}
\put(37.984, 68.462){\circle*{1.0}}
\put(38.112, 67.876){\circle*{1.0}}
\put(38.240, 67.290){\circle*{1.0}}
\put(38.240, 67.290){\line(-1,0){5}}
\put(38.368, 66.704){\circle*{1.0}}
\put(38.368, 66.704){\line(-1,0){5}}
\put(38.496, 66.118){\circle*{1.0}}
\put(38.496, 66.118){\line(-1,0){5}}
\put(38.624, 65.531){\circle*{1.0}}
\put(38.624, 65.531){\line(-1,0){5}}
\put(38.713, 64.938){\circle*{1.0}}
\put(38.802, 64.345){\circle*{1.0}}
\put(38.890, 63.751){\circle*{1.0}}
\put(38.979, 63.158){\circle*{1.0}}
\put(39.068, 62.564){\circle*{1.0}}
\put(39.068, 62.564){\line(-1,0){5}}
\put(39.157, 61.971){\circle*{1.0}}
\put(39.157, 61.971){\line(-1,0){5}}
\put(39.245, 61.378){\circle*{1.0}}
\put(39.245, 61.378){\line(-1,0){5}}
\put(39.334, 60.784){\circle*{1.0}}
\put(39.334, 60.784){\line(-1,0){5}}
\put(39.389, 60.187){\circle*{1.0}}
\put(39.444, 59.589){\circle*{1.0}}
\put(39.500, 58.992){\circle*{1.0}}
\put(39.555, 58.394){\circle*{1.0}}
\put(39.610, 57.797){\circle*{1.0}}
\put(39.610, 57.797){\line(-1,0){5}}
\put(39.665, 57.199){\circle*{1.0}}
\put(39.665, 57.199){\line(-1,0){5}}
\put(39.720, 56.602){\circle*{1.0}}
\put(39.720, 56.602){\line(-1,0){5}}
\put(39.775, 56.005){\circle*{1.0}}
\put(39.775, 56.005){\line(-1,0){5}}
\put(39.830, 55.407){\circle*{1.0}}
\put(39.849, 54.807){\circle*{1.0}}
\put(39.868, 54.208){\circle*{1.0}}
\put(39.887, 53.608){\circle*{1.0}}
\put(39.906, 53.008){\circle*{1.0}}
\put(39.906, 53.008){\line(-1,0){5}}
\put(39.924, 52.409){\circle*{1.0}}
\put(39.924, 52.409){\line(-1,0){5}}
\put(39.943, 51.809){\circle*{1.0}}
\put(39.943, 51.809){\line(-1,0){5}}
\put(39.962, 51.209){\circle*{1.0}}
\put(39.962, 51.209){\line(-1,0){5}}
\put(39.981, 50.609){\circle*{1.0}}
\put(40.000, 50.010){\circle*{1.0}}
\put(39.982, 49.410){\circle*{1.0}}
\put(39.965, 48.810){\circle*{1.0}}
\put(39.948, 48.210){\circle*{1.0}}
\put(39.948, 48.210){\line(-1,0){5}}
\put(39.931, 47.611){\circle*{1.0}}
\put(39.931, 47.611){\line(-1,0){5}}
\put(39.913, 47.011){\circle*{1.0}}
\put(39.913, 47.011){\line(-1,0){5}}
\put(39.896, 46.411){\circle*{1.0}}
\put(39.896, 46.411){\line(-1,0){5}}
\put(39.879, 45.811){\circle*{1.0}}
\put(39.862, 45.212){\circle*{1.0}}
\put(39.812, 44.614){\circle*{1.0}}
\put(39.763, 44.016){\circle*{1.0}}
\put(39.713, 43.418){\circle*{1.0}}
\put(39.713, 43.418){\line(-1,0){5}}
\put(39.663, 42.820){\circle*{1.0}}
\put(39.663, 42.820){\line(-1,0){5}}
\put(39.614, 42.222){\circle*{1.0}}
\put(39.614, 42.222){\line(-1,0){5}}
\put(39.564, 41.624){\circle*{1.0}}
\put(39.564, 41.624){\line(-1,0){5}}
\put(39.515, 41.026){\circle*{1.0}}
\put(39.465, 40.428){\circle*{1.0}}
\put(39.415, 39.830){\circle*{1.0}}
\put(39.329, 39.236){\circle*{1.0}}
\put(39.243, 38.643){\circle*{1.0}}
\put(39.243, 38.643){\line(-1,0){5}}
\put(39.156, 38.049){\circle*{1.0}}
\put(39.156, 38.049){\line(-1,0){5}}
\put(39.070, 37.455){\circle*{1.0}}
\put(39.070, 37.455){\line(-1,0){5}}
\put(38.984, 36.861){\circle*{1.0}}
\put(38.984, 36.861){\line(-1,0){5}}
\put(38.898, 36.268){\circle*{1.0}}
\put(38.811, 35.674){\circle*{1.0}}
\put(38.725, 35.080){\circle*{1.0}}
\put(38.605, 34.492){\circle*{1.0}}
\put(38.484, 33.905){\circle*{1.0}}
\put(38.484, 33.905){\line(-1,0){5}}
\put(38.363, 33.317){\circle*{1.0}}
\put(38.363, 33.317){\line(-1,0){5}}
\put(38.243, 32.729){\circle*{1.0}}
\put(38.243, 32.729){\line(-1,0){5}}
\put(38.122, 32.141){\circle*{1.0}}
\put(38.122, 32.141){\line(-1,0){5}}
\put(38.002, 31.554){\circle*{1.0}}
\put(37.881, 30.966){\circle*{1.0}}
\put(37.761, 30.378){\circle*{1.0}}
\put(37.640, 29.790){\circle*{1.0}}
\put(37.479, 29.212){\circle*{1.0}}
\put(37.479, 29.212){\line(-1,0){5}}
\put(37.319, 28.634){\circle*{1.0}}
\put(37.319, 28.634){\line(-1,0){5}}
\put(37.158, 28.056){\circle*{1.0}}
\put(37.158, 28.056){\line(-1,0){5}}
\put(36.997, 27.478){\circle*{1.0}}
\put(36.997, 27.478){\line(-1,0){5}}
\put(36.836, 26.900){\circle*{1.0}}
\put(36.676, 26.322){\circle*{1.0}}
\put(36.515, 25.744){\circle*{1.0}}
\put(36.354, 25.166){\circle*{1.0}}
\put(36.155, 24.600){\circle*{1.0}}
\put(36.155, 24.600){\line(-1,0){5}}
\put(35.956, 24.034){\circle*{1.0}}
\put(35.956, 24.034){\line(-1,0){5}}
\put(35.756, 23.468){\circle*{1.0}}
\put(35.756, 23.468){\line(-1,0){5}}
\put(35.557, 22.902){\circle*{1.0}}
\put(35.557, 22.902){\line(-1,0){5}}
\put(35.358, 22.336){\circle*{1.0}}
\put(35.158, 21.770){\circle*{1.0}}
\put(34.959, 21.204){\circle*{1.0}}
\put(34.723, 20.653){\circle*{1.0}}
\put(34.487, 20.101){\circle*{1.0}}
\put(34.487, 20.101){\line(-1,0){5}}
\put(34.251, 19.549){\circle*{1.0}}
\put(34.251, 19.549){\line(-1,0){5}}
\put(34.015, 18.998){\circle*{1.0}}
\put(34.015, 18.998){\line(-1,0){5}}
\put(33.778, 18.446){\circle*{1.0}}
\put(33.778, 18.446){\line(-1,0){5}}
\put(33.542, 17.895){\circle*{1.0}}
\put(33.306, 17.343){\circle*{1.0}}
\put(33.070, 16.792){\circle*{1.0}}
\put(32.789, 16.261){\circle*{1.0}}
\put(32.508, 15.731){\circle*{1.0}}
\put(32.508, 15.731){\line(-1,0){5}}
\put(32.227, 15.201){\circle*{1.0}}
\put(32.227, 15.201){\line(-1,0){5}}
\put(31.946, 14.671){\circle*{1.0}}
\put(31.946, 14.671){\line(-1,0){5}}
\put(31.666, 14.141){\circle*{1.0}}
\put(31.666, 14.141){\line(-1,0){5}}
\put(31.385, 13.610){\circle*{1.0}}
\put(31.104, 13.080){\circle*{1.0}}
\put(30.779, 12.576){\circle*{1.0}}

\put(65,50){\vector(1,0){50}} 
\put(65,59){\vector(4,1){50}} 
\put(65,41){\vector(4,-1){50}} 
\multiput(77,93)(2,-2){3}{\vector(-1,-1){20}}
\put(61,80){\makebox(20,8)[l]{$\bf v$}}
\put(98,50){\circle*{1.5}}
\put(98,52.3){\circle*{1.5}}
\put(97.8,54.6){\circle*{1.5}}
\put(97.6,56.9){\circle*{1.5}}
\put(97.3,59.2){\circle*{1.5}}
\put(96.9,61.5){\circle*{1.5}}
\put(96.4,63.7){\circle*{1.5}}
\put(95.8,65.8){\circle*{1.5}}
\end{picture}
\begin{minipage}{4.0in}

\vskip-0.0cm
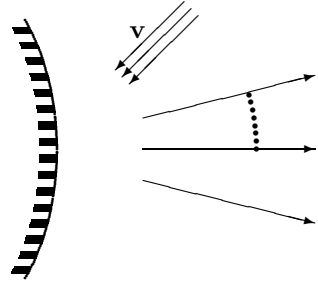
Fig.~2: Schematic of reflected particles from a solid surface. It is 
interesting to note that whether or not the involved collisions are 
elastic will not alter this picture significantly.
\end{minipage}

\begin{picture}(100,103)
\hspace{-2.3cm}
\multiput(15,19)(90,0){2}{\vector(1,0){80}} 
\multiput(20,15)(90,0){2}{\vector(0,1){60}} 
\put(45,40){\framebox(20,20){}} 
\multiput(45,40)(1.3,1.7){4}{\circle*{1.5}} 
\multiput(50.2,46.8)(1.5,1.5){6}{\circle*{1.5}} 
\multiput(59.2,55.8)(1.7,1.3){4}{\circle*{1.5}} 
\multiput(135,40)(20,20){2}{\line(1,0){20}} 
\multiput(135,40)(20,0){2}{\line(1,1){20}} 
\multiput(135,40)(3,1.7){4}{\circle*{1.5}} 
\multiput(147,47.2)(3,1.5){6}{\circle*{1.5}} 
\multiput(165,56.2)(3,1.2){4}{\circle*{1.5}} 
\put(50,5){\makebox(20,8)[l]{\bf (a)}}
\put(140,5){\makebox(20,8)[l]{\bf (b)}}
\multiput(89,15)(90,0){2}{\makebox(20,8)[c]{$x$}}
\multiput(10,77)(90,0){2}{\makebox(20,8)[c]{$v_x$}}
\put(46,32){\makebox(20,8)[c]{\small $\Delta x$}}
\put(27,47){\makebox(20,8)[c]{\small $\Delta v_x$}}
\end{picture}
\begin{minipage}{4.0in}
\vskip0.2cm
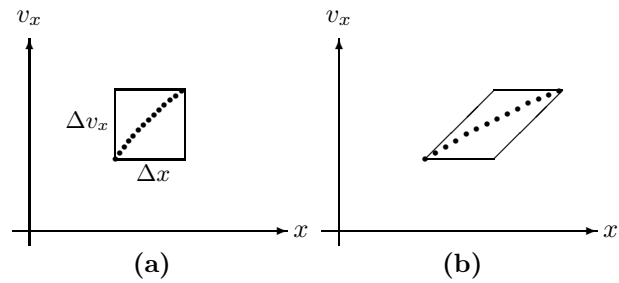
Fig.~3: Schematic of how the particles marked in Fig.~2 spread in the 
$x-v_x$ space. (a) 
These particles are distributed along one diagonal of the rectangle 
$\Delta x\Delta v_x$ at $t=0$, and (b) the diagonal is  stretched at $t=T$.
\end{minipage}

\begin{picture}(100,100)
\hspace{-1.0cm}
\put(30.797, 87.459){\circle*{1.0}}
\put(31.083, 86.931){\circle*{1.0}}
\put(31.369, 86.404){\circle*{1.0}}
\put(31.656, 85.877){\circle*{1.0}}
\put(31.942, 85.349){\circle*{1.0}}
\put(32.228, 84.822){\circle*{1.0}}
\put(32.514, 84.294){\circle*{1.0}}
\put(32.800, 83.767){\circle*{1.0}}
\put(33.044, 83.219){\circle*{1.0}}
\put(33.288, 82.671){\circle*{1.0}}
\put(33.532, 82.122){\circle*{1.0}}
\put(33.776, 81.574){\circle*{1.0}}
\put(34.020, 81.026){\circle*{1.0}}
\put(34.263, 80.478){\circle*{1.0}}
\put(34.507, 79.930){\circle*{1.0}}
\put(34.751, 79.381){\circle*{1.0}}
\put(34.951, 78.816){\circle*{1.0}}
\put(35.152, 78.250){\circle*{1.0}}
\put(35.352, 77.685){\circle*{1.0}}
\put(35.552, 77.119){\circle*{1.0}}
\put(35.753, 76.553){\circle*{1.0}}
\put(35.953, 75.988){\circle*{1.0}}
\put(36.153, 75.422){\circle*{1.0}}
\put(36.318, 74.845){\circle*{1.0}}
\put(36.483, 74.269){\circle*{1.0}}
\put(36.648, 73.692){\circle*{1.0}}
\put(36.813, 73.115){\circle*{1.0}}
\put(36.978, 72.538){\circle*{1.0}}
\put(37.143, 71.961){\circle*{1.0}}
\put(37.308, 71.384){\circle*{1.0}}
\put(37.472, 70.807){\circle*{1.0}}
\put(37.600, 70.221){\circle*{1.0}}
\put(37.728, 69.635){\circle*{1.0}}
\put(37.856, 69.049){\circle*{1.0}}
\put(37.984, 68.462){\circle*{1.0}}
\put(38.112, 67.876){\circle*{1.0}}
\put(38.240, 67.290){\circle*{1.0}}
\put(38.368, 66.704){\circle*{1.0}}
\put(38.496, 66.118){\circle*{1.0}}
\put(38.624, 65.531){\circle*{1.0}}
\put(38.713, 64.938){\circle*{1.0}}
\put(38.802, 64.345){\circle*{1.0}}
\put(38.890, 63.751){\circle*{1.0}}
\put(38.979, 63.158){\circle*{1.0}}
\put(39.068, 62.564){\circle*{1.0}}
\put(39.157, 61.971){\circle*{1.0}}
\put(39.245, 61.378){\circle*{1.0}}
\put(39.334, 60.784){\circle*{1.0}}
\put(39.389, 60.187){\circle*{1.0}}
\put(39.444, 59.589){\circle*{1.0}}
\put(39.500, 58.992){\circle*{1.0}}
\put(39.555, 58.394){\circle*{1.0}}
\put(39.610, 57.797){\circle*{1.0}}
\put(39.665, 57.199){\circle*{1.0}}
\put(39.720, 56.602){\circle*{1.0}}
\put(39.775, 56.005){\circle*{1.0}}
\put(39.830, 55.407){\circle*{1.0}}
\put(39.849, 54.807){\circle*{1.0}}
\put(39.868, 54.208){\circle*{1.0}}
\put(39.887, 53.608){\circle*{1.0}}
\put(39.906, 53.008){\circle*{1.0}}
\put(39.924, 52.409){\circle*{1.0}}
\put(39.943, 51.809){\circle*{1.0}}
\put(39.962, 51.209){\circle*{1.0}}
\put(39.981, 50.609){\circle*{1.0}}
\put(40.000, 50.010){\circle*{1.0}}
\put(39.982, 49.410){\circle*{1.0}}
\put(39.965, 48.810){\circle*{1.0}}
\put(39.948, 48.210){\circle*{1.0}}
\put(39.931, 47.611){\circle*{1.0}}
\put(39.913, 47.011){\circle*{1.0}}
\put(39.896, 46.411){\circle*{1.0}}
\put(39.879, 45.811){\circle*{1.0}}
\put(39.862, 45.212){\circle*{1.0}}
\put(39.812, 44.614){\circle*{1.0}}
\put(39.763, 44.016){\circle*{1.0}}
\put(39.713, 43.418){\circle*{1.0}}
\put(39.663, 42.820){\circle*{1.0}}
\put(39.614, 42.222){\circle*{1.0}}
\put(39.564, 41.624){\circle*{1.0}}
\put(39.515, 41.026){\circle*{1.0}}
\put(39.465, 40.428){\circle*{1.0}}
\put(39.415, 39.830){\circle*{1.0}}
\put(39.329, 39.236){\circle*{1.0}}
\put(39.243, 38.643){\circle*{1.0}}
\put(39.156, 38.049){\circle*{1.0}}
\put(39.070, 37.455){\circle*{1.0}}
\put(38.984, 36.861){\circle*{1.0}}
\put(38.898, 36.268){\circle*{1.0}}
\put(38.811, 35.674){\circle*{1.0}}
\put(38.725, 35.080){\circle*{1.0}}
\put(38.605, 34.492){\circle*{1.0}}
\put(38.484, 33.905){\circle*{1.0}}
\put(38.363, 33.317){\circle*{1.0}}
\put(38.243, 32.729){\circle*{1.0}}
\put(38.122, 32.141){\circle*{1.0}}
\put(38.002, 31.554){\circle*{1.0}}
\put(37.881, 30.966){\circle*{1.0}}
\put(37.761, 30.378){\circle*{1.0}}
\put(37.640, 29.790){\circle*{1.0}}
\put(37.479, 29.212){\circle*{1.0}}
\put(37.319, 28.634){\circle*{1.0}}
\put(37.158, 28.056){\circle*{1.0}}
\put(36.997, 27.478){\circle*{1.0}}
\put(36.836, 26.900){\circle*{1.0}}
\put(36.676, 26.322){\circle*{1.0}}
\put(36.515, 25.744){\circle*{1.0}}
\put(36.354, 25.166){\circle*{1.0}}
\put(36.155, 24.600){\circle*{1.0}}
\put(35.956, 24.034){\circle*{1.0}}
\put(35.756, 23.468){\circle*{1.0}}
\put(35.557, 22.902){\circle*{1.0}}
\put(35.358, 22.336){\circle*{1.0}}
\put(35.158, 21.770){\circle*{1.0}}
\put(34.959, 21.204){\circle*{1.0}}
\put(34.723, 20.653){\circle*{1.0}}
\put(34.487, 20.101){\circle*{1.0}}
\put(34.251, 19.549){\circle*{1.0}}
\put(34.015, 18.998){\circle*{1.0}}
\put(33.778, 18.446){\circle*{1.0}}
\put(33.542, 17.895){\circle*{1.0}}
\put(33.306, 17.343){\circle*{1.0}}
\put(33.070, 16.792){\circle*{1.0}}
\put(32.789, 16.261){\circle*{1.0}}
\put(32.508, 15.731){\circle*{1.0}}
\put(32.227, 15.201){\circle*{1.0}}
\put(31.946, 14.671){\circle*{1.0}}
\put(31.666, 14.141){\circle*{1.0}}
\put(65,50){\vector(1,0){50}} 
\put(65,57){\vector(4,1){50}} 
\put(65,43){\vector(4,-1){50}} 
\put(80,90){\vector(-1,-1){15}}
\put(78.6,91.4){\vector(-1,-1){15}}
\put(60,95){\vector(-1,-2){10}}
\put(58.4,95.8){\vector(-1,-2){10}}
\put(101,54){\makebox(20,8)[l]{${\bf r}$}}
\put(16,42){\makebox(20,8)[r]{$(\Delta S)_i$}}
\put(100,54){\circle*{2}}
\put(38.5,50){\oval(3,5)}
\end{picture}
\begin{minipage}{4.0in}
\vskip0.3cm
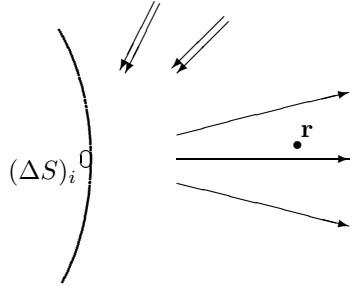
Fig.~4: Particles reflected from a small surface element of realistic 
boundary.
\end{minipage}

\begin{picture}(80,100)
\hspace{-0.8cm}
\put(57,53.5){\vector(2,1){20}} 
\put(57,46.5){\vector(2,-1){20}} 
\multiput(50,50)(2,1){4}{\circle*{0.5}} 
\multiput(50,50)(2,-1){4}{\circle*{0.5}} 
\put(39,58){\makebox(20,8)[c]{$d{\bf r}d{\bf v}$}}
\put(70,62){\makebox(20,8)[c]{${\bf v}$}}
\put(70,37){\makebox(20,8)[c]{${\bf v}^\prime$}}
\put(50,50){\circle{15}}
\put(50,50){\line(-2,-1){27.5}}
\end{picture}

\begin{minipage}{4.0in}
\vskip-0.2cm
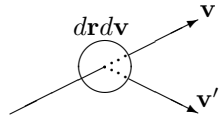
Fig.~5: A particle involving a collision in a phase volume element.
\end{minipage}

\begin{picture}(100,95)
\hspace{-0.1cm}
\multiput(0,51)(0.8,1.6){2}{\vector(2,-1){37.5}}
\put(54,25.5){\vector(2,-1){37.5}}
\put(54.5,27){\vector(3,-1){40}}
\put(53,24){\vector(3,-2){34.5}}
\multiput(0,9)(0.8,-1.6){2}{\vector(2,1){37.5}}
\put(54,34.5){\vector(2,1){37.5}}
\put(53,35.5){\vector(3,2){34.5}}
\put(54.5,33){\vector(3,1){40}}
\put(10,8){\makebox(20,8)[c]{${\bf v}_1^\prime$}}
\put(10,44){\makebox(20,8)[c]{${\bf v}^\prime$}}
\put(70,18){\makebox(20,8)[c]{${\bf v}_1$}}
\put(68,52){\makebox(20,8)[c]{${\bf v}$}}
\end{picture}
\begin{minipage}{4.0in}
\vskip1.0cm
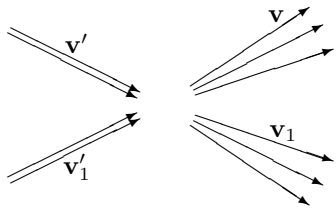
Fig.~6: Two beams of particles collide with each other and produce 
particles with velocity ${\bf v}$.
\end{minipage}
\end{center}

\end{document}